\documentclass[runningheads]{llncs}
\usepackage{graphicx}

\usepackage{subcaption}
\usepackage{amsmath}
\usepackage{amssymb}
\usepackage{booktabs}
\usepackage[pagebackref,breaklinks,colorlinks]{hyperref}
\usepackage[capitalize]{cleveref}
\crefname{section}{Sec.}{Secs.}
\Crefname{section}{Section}{Sections}
\Crefname{table}{Table}{Tables}
\crefname{table}{Tab.}{Tabs.}
\begin{document}
\title{Is visual explanation with Grad-CAM more reliable for deeper neural networks? a case study with automatic pneumothorax diagnosis}
%
%
\author{Zirui Qiu\inst{1} \and
Hassan Rivaz\inst{2}\and
Yiming Xiao\inst{1}}

\authorrunning{Z.Qiu et al.}
%
\institute{Department of Computer Science and Software Engineering, Concordia University, Montreal, Canada \and
Department of Electrical and Computer Engineering, Concordia University, Montreal, Canada
}

\maketitle              
\begin{abstract}
While deep learning techniques have provided the state-of-the-art performance in various clinical tasks, explainability regarding their decision-making process can greatly enhance the credence of these methods for safer and quicker clinical adoption. With high flexibility, Gradient-weighted Class Activation Mapping (Grad-CAM) has been widely adopted to offer intuitive visual interpretation of various deep learning models' reasoning processes in computer-assisted diagnosis. However, despite the popularity of the technique, there is still a lack of systematic study on Grad-CAM's performance on different deep learning architectures. In this study, we investigate its robustness and effectiveness across different  popular deep learning models, with a focus on the impact of the networks' depths and architecture types, by using a case study of automatic pneumothorax diagnosis in X-ray scans. Our results show that deeper neural networks do not necessarily contribute to a strong improvement of pneumothorax diagnosis accuracy, and the effectiveness of GradCAM also varies among different network architectures.

\end{abstract}
\section{Introduction}
With rapid development, deep learning (DL) techniques have become the state-of-the-art in many vision applications, such as computer-assisted diagnosis. Although initially proposed for natural image processing, staple Convolutional Neural Networks (CNNs), such as VGG and ResNet architectures, have become ubiquitous backbones in computer-assisted radiological applications due to their robustness and flexibility to capture task-specific, complex image features. Furthermore, the more recent Vision Transformer (ViT), a new class of DL architecture that leverage the self-attention mechanism to encode long-range contextual information is attracting great attention in medical image processing, with evidence showing superior performance than the more traditional CNNs \cite{chen2021transunet}. Although these DL algorithms can offer excellent accuracy, one major challenge that hinders their wide adoption in clinical practice is the lack of transparency and interpretability in their decision-making process. So far, various explainable AI (XAI) methods have been proposed \cite{arrieta2020explainable}, and among these, direct visualization of the saliency/activation maps has gained high popularity, likely due to their intuitiveness for fast uptake in clinical applications. With high flexibility and ease of implementation for different DL architectures, the Gradient-weighted Class Activation Mapping (Grad-CAM) technique \cite{selvaraju2017grad}, which provides visual explanation as heatmaps with respect to class-wise decision has been applied widely in many computer-assisted diagnostic and surgical vision applications. However, in almost all previous investigations, Grad-CAM outcomes are only demonstrated qualitatively. To the best of our knowledge, the impacts of different deep learning architectures and sizes on the robustness and effectiveness of Grad-CAM have not been investigated, but are important for the research community. 

To address the mentioned knowledge gap, we benchmarked the performance of Grad-CAM on DL models of three types of popular architectures, including VGG, ResNet, and ViT, with varying network depths/sizes of each one. We explored the impacts of DL architectures on GradCAM by using pneumothorax diagnosis from chest X-ray images as a case study. Pneumothorax is a condition that is characterized by the accumulation of air in the pleural space, and can lead to lung collapse, posing a significant risk to patient health if not promptly diagnosed and treated. As the radiological features of pneumothorax (i.e., air invasion) can be subtle to spot in X-ray scans, the task provides an excellent case to examine the characteristics of different DL models and Grad-CAM visualization. In summary, our work has two main contributions. \textbf{First}, we conducted a comprehensive evaluation of popular DL models including CNNs and Transformers for pneumothorax diagnosis. \textbf{Second}, we systematically compared the effectiveness of visual explanation using Grad-CAM across these staple DL models both qualitatively and quantitatively. Here, we analyzed the impact of network architecture choices on diagnostic accuracy and effectiveness of Grad-CAM results.

\vspace{-0.1cm}
\section{Related Works}
The great accessibility of public chest X-ray datasets has allowed a large amount works \cite{mijwil2021implementation} on the diagnosis and segmentation of lung diseases using deep learning algorithms, with a comprehensive review provided by Calli et al. \cite{ccalli2021deep}. So far, many previous reports adopted popular DL models that were first designed for natural image processing. For example, Tian et al. \cite{tian2022deep} leveraged ResNet and VGG models with multi-instance transfer learning for pneumothorax classification. Wollek at al. \cite{wollek2022attention} employed the Vision Transformer to perform automatic diagnosis for multiple lung conditions on chest X-rays. To incorporate visual explanation for DL-based pneumothorax diagnosis, many have adopted Grad-CAM and its variants \cite{wollek2022attention,yuan2023human} for both CNNs and ViTs. Yuan et al. \cite{yuan2023human} proposed a human-guided design to enhance the performance of Saliency Map, Grad-CAM, and Integrated Gradients in visual interpretability of pneumothorax diagnosis using CNNs. Most recently, Sun et al. \cite{sun2023inherently} proposed the Attri-Net, which employed Residual blocks and multi-label explanations that align with clinical knowledge for improved visual explanation in chest X-ray classification.  

\section{Material and Methodology}
\label{sec:Methodology}


\subsection{Deep Learning Model Architectures}

Our study explored a selection of staple deep learning architectures, including VGG, ResNet and ViT, which are widely used in both natural and medical images. Specifically, the VGG models \cite{simonyan2014very} are characterized by their multiple 3x3 convolution layers, and for the study, we included VGG-16 and VGG-19. The ResNet models \cite{he2016deep} leverage skip connections to enhance residual learning and training stability. Here, we incorporated ResNet18, ResNet34, ResNet50, and ResNet101, which comprise 18, 34, 50, and 101 layers, respectively. Lastly, the Vision Transformers initially proposed by Dosovitskiy et al.\cite{dosovitskiy2020image} treat images as sequences of patches/tokens to model their long-range dependencies without the use of convolution operations. To test the influence of network sizes, we adopted the $ViT\_small$ and $ViT\_base$ variants, both with 12 layers but differing in input features, and the $ViT\_large$ variant with 24 layers. All these models were pretrained on ImageNet-1K and subsequently fine-tuned for the task of Pneumothorax vs. Healthy classification using the curated public dataset. 

\subsection{Grad-CAM Visualization}
To infer the decision-making process of DL models, the Grad-CAM technique \cite{selvaraju2017grad} creates a heatmap by computing the gradients of the target class score with respect to the feature maps of the last convolutional layer. Specifically, for VGG16 and VGG19, we applied Grad-CAM to the last convolution layer. For ResNet models, we targeted the final bottleneck layer and in Vision Transformer variants, the technique was applied to the final block layer before the classification token is processed. Ideally, an effective visual guidance should provide high accuracy (i.e., correct identification of the region of interest by high values in the heatmap) and specificity (i.e., tight bound around the region of interest). Note that for each Grad-CAM heatmap, the value is normalized to [0,1].
\subsection{Dataset Preprocessing and Experimental Setup}
For this study, we used the SIIM-ACR Pneumothorax Segmentation dataset from Kaggle \footnote{SIIM-ACR Pneumothorax Segmentation:\url{https://www.kaggle.com/competitions/siim-acr-pneumothorax-segmentation/data}}. It contains chest X-Ray images of 9,000 healthy controls and 3,600 patients with pneumothorax. In addition, regions related to pneumothorax were manually segmented for 3576 patients. For our experiments, we created a balanced subset of 7,200 cases (50\% with pneumothorax) from the original dataset. From the curated data collection, we divided the cases into 7,000 for training, 1,000 for validation, and 1,000 for testing while balancing the health vs. pneumothorax ratio in each set. For DL model training and testing, each image was processed using Contrast Limited Adaptive Histogram Equalization (CLAHE) and normlaized with z-transform. In addition, all images were re-scaled to the common dimension of 224$\times$224 pixels. In terms of training, the VGG and ResNet models utilized the cross-entropy loss function with the Adam optimizer, and were trained at a learning rate of 1e-4 for 50 epochs. The ViT models employed a cross-entropy loss function and the Stochastic Gradient Descent (SGD) method for optimization with a learning rate of 1e-4. They were trained for 300 epochs. Finally,  to boost our model's performance and mitigate overfitting, we used data augmentations in training, including the addition of random Gaussian noise, rotations (up to 10 degrees), horizontal flips, and brightness/contrast shifts.

\subsection{Evaluation Metrics}
To evaluate the performance of different DL models in pneumothorax diagnosis, we assessed the accuracy, precision, recall, and area under the curve (AUC) metrics. In terms of assessing the effectiveness of the Grad-CAM results for each model, we propose to use two different measures. First, we compute the difference between the means of the Grad-CAM heatmap values within and outside the ground truth pneumothorax segmentation, and refer to this metric as $Diff_{GradCAM}$. We hypothesize that an effective Grad-CAM visualization should generate a high positive $Diff_{GradCAM}$ because the ideal heatmap should accumulate high values primarily within the pathological region (i.e., air invasion in the pleural space). The scores from the models were further compared by a one-way ANOVA test and Tukey's post-hoc analysis, and a p-value $<0.05$ indicated a statistically significant difference. Second, we compute the Effective Heat Ratio (EHR) \cite{wollek2022attention}, which is the ratio between the thresholded area of the Grad-CAM heatmap within the ground truth segmentation and the total threshold area. The thresholds were computed in equidistant steps, and the Area Under the Curve (AUC) is calculated over all EHRs and the associated threshold values to assess the quality of Grad-CAM results. Both metrics reflect the accuracy and specificity of visual explanation for the networks.

\section{Results}

\subsection{Pneumothorax Diagnosis Performance}
The performance of pneumothorax diagnosis for all DL models is listed in Table 1. In terms of accuracy, ResNet models offered the best results, particularly with ResNet50 at 88.20$\%$, and the ViT and VGG ranked the second and the last. The similar trend held for precision and AUC. However, as for recall, the obtained scores were similar across different architecture types. When looking into different network sizes for each architecture type, the results showed that deeper neural networks did not necessarily produce superior diagnostic performance. Specifically, the two popular VGG models didn't result in large discrepancy in accuracy, recall, and AUC while VGG16 has better precision. For the ResNet models, ResNet18, ResNet34 and ResNet101 had similar performance, with the accuracy, recall, and AUC peaked slightly at ResNet50. This likely means that ResNet18 has enough representation power to perform the classification, and therefore deeper networks do not improve the results.  Finally, for the ViTs, $ViT\_base$ resulted in better performance than the small and large versions, with slight performance deterioration for $ViT\_large$ . 

\begin{table}[h]
\centering
\caption{Pneumothorax diagnosis performance across all DL models}
\begin{tabular}{|l|l|l|l|l|}
\hline
Model & Accuracy & Precision & Recall & AUC \\ \hline
VGG16 & 83.70\% & 0.8224 & 0.8800 & 0.9069 \\ \hline
VGG19 & 83.20\% & 0.7747 & 0.9080 & 0.9098 \\ \hline
ResNet18  & 87.60\%  & 0.8821    & 0.8680 & 0.9420  \\ \hline
ResNet34  & 87.20\%  & 0.8758    & 0.8600 & 0.9417  \\ \hline
ResNet50  & 88.20\%  & 0.8802    & 0.8820 & 0.9450  \\ \hline
ResNet101 & 87.60\%  & 0.8798    & 0.8780 & 0.9434  \\ \hline
ViT Small & 85.80\%  & 0.8137    & 0.8820 & 0.9302  \\ \hline
ViT Base  & 86.20\%  & 0.8602    & 0.8740 & 0.9356  \\ \hline
ViT Large & 84.40\%  & 0.8356    & 0.8640 & 0.9197  \\ \hline
\end{tabular}

\label{tab:model_performance}
\end{table}

\subsection{Qualitative Grad-CAM Evaluation}
We present the Grad-CAM heatmaps for three patient cases across all tested DL models in Fig. 1, with the ground truth pneumothorax segmentation overlaid in the X-ray scans. In the presented cases, all the heatmaps correctly indicated the side of the lungs affected by the disease. However, the overlap with the pneumothorax segmentation varied. The VGG16 and VGG19 models pinpointed the pneumothorax areas for the first two cases while VGG19 failed to do so for the third case. Note that both models presented a secondary region. The ResNet18, 34, and 50 models successfully highlighted the problematic area, with the heatmap of ResNet18 slightly off-center, while the ResNet101 model showed activation in two regions. In comparison, ViT models exhibited more dispersed Grad-CAM patterns than the CNNs, and the amount of unrelated areas increased with the model size.

\begin{figure}[!htbp]
\centering
\includegraphics[width=1.0\textwidth]{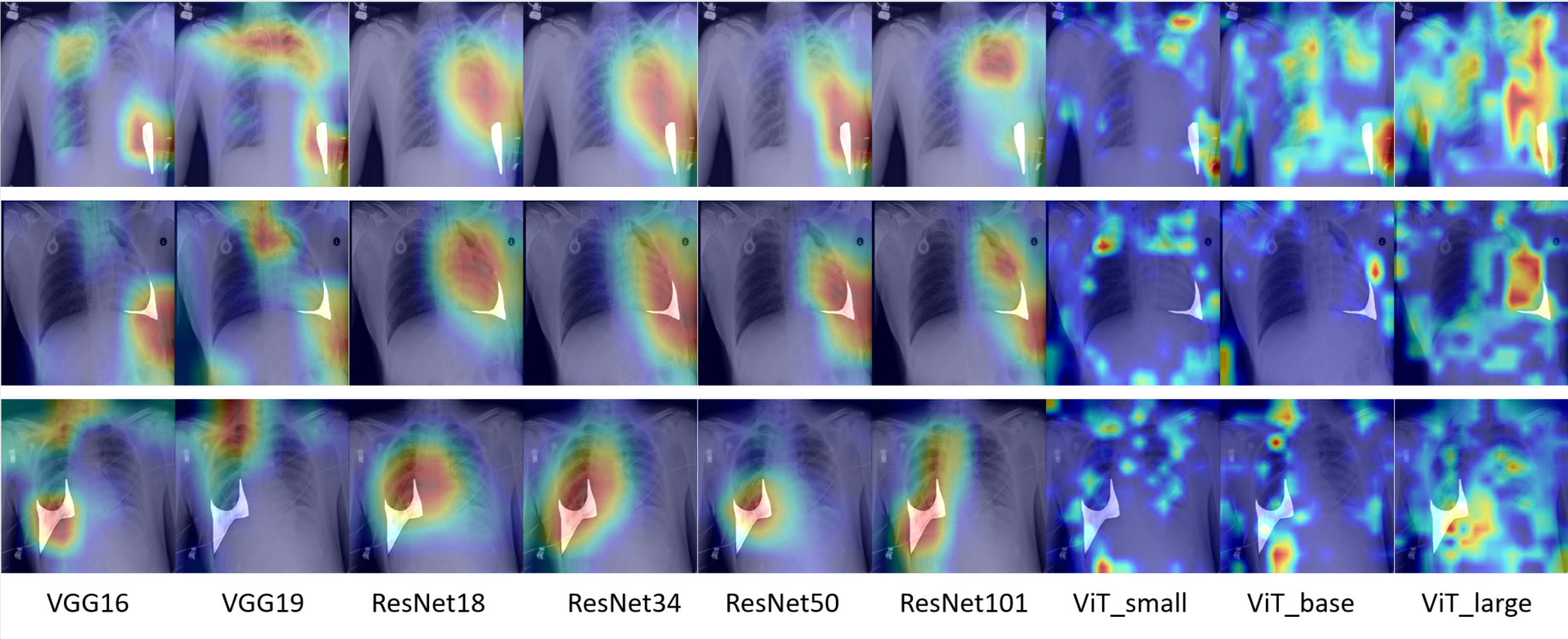}
\caption{\label{fig:ViT} Demonstration of Grad-CAM results across different deep learning models for three patients X-ray images (one patient per row), with the manual pneumothorax segmentation overlaid in white color.}
\end{figure}

\subsection{Quantitative Grad-CAM Evaluation}

\begin{table}[h]
\centering
\caption{Difference of mean Grad-CAM values within and outside the manual pneumothorax segmentation ($Diff_{GradCAM}$).}
\begin{tabular}{lrrrrrrrrr}
\toprule
{} &   VGG16 &   VGG19 &  ResNet18 &  ResNet34 &  ResNet50 &  ResNet101 &  ViT\_small &  ViT\_base &  ViT\_large \\
\midrule
mean  &    0.186 &    0.166 &     0.162 &     0.133 &     0.142 &     0.183 &     0.052 &     0.081 &     0.051 \\
std   &    0.212 &    0.220 &     0.262 &     0.273 &     0.251 &     0.265 &     0.143 &     0.166 &     0.164 \\
min   &   -0.213 &   -0.365 &    -0.366 &    -0.452 &    -0.287 &    -0.320 &    -0.251 &    -0.233 &    -0.319 \\
max   &    0.773 &    0.715 &     0.787 &     0.782 &     0.801 &     0.780 &     0.652 &     0.786 &     0.685 \\
\bottomrule
\end{tabular}

\label{tab:my_label}
\end{table}

For the Grad-CAM heatmap of each tested model, we computed the $Diff_{GradCAM}$ and EHR AUC metrics across all test cases to gauge their effectiveness of visually interpreting the decision-making process of DL algorithms. Here, $Diff_{GradCAM}$ and EHR AUC are reported in Table2 and Fig. 2, respectively. For $Diff_{GradCAM}$, the ANOVA test showed a group-wise difference (p$<$0.001). In general, the CNNs offered better results than the ViT models, despite ViTs' good pneumothorax diagnosis accuracy. VGG16 and ResNet101 ranked the best (p$<$0.05), but the associated standard deviations of CNN models were also higher. Within each architecture type,the scores for VGG16 and VGG19 were similar (p$>$0.05), the ViT models also don't differ significantly (p$>$0.05), and ResNet101's score was significantly higher than ResNet34 (p$<$0.05). Between VGG and ResNet models, comparisons between ResNet18, ResNet50, VGG16, and VGG19 did not yield any significant differences (p$>$0.05). Among all the tested models, ResNet50 achieved the highest max score and ResNet34 had the lowest min score. As for EHR AUC, similar to the case of $Diff_{GradCAM}$, the CNN models also outperformed the ViT ones, with ResNet101 leading the scores at 0.0319 and VGG16 ranking the second at 0.0243. Compared with VGG16, the EHR AUC score of VGG19 was very similar. Among the ResNet variants, ResNet18, ResNet34, and ResNet50 scored similary in the range of 0.21$\sim$0.23, generally lower than those of VGG models. With the deepest architecture among the ResNet models, ResNet101 had a large increase of the EHR AUC metric even though its pneumothorax diagnosis accuracy was similar to the rest. For the ViT models, the EHR AUC improved gradually with the increasing size of the architecture, ranging from 0.0145 to 0.0171.

\begin{figure}[!htbp]
\centering
\includegraphics[width=0.9\textwidth]{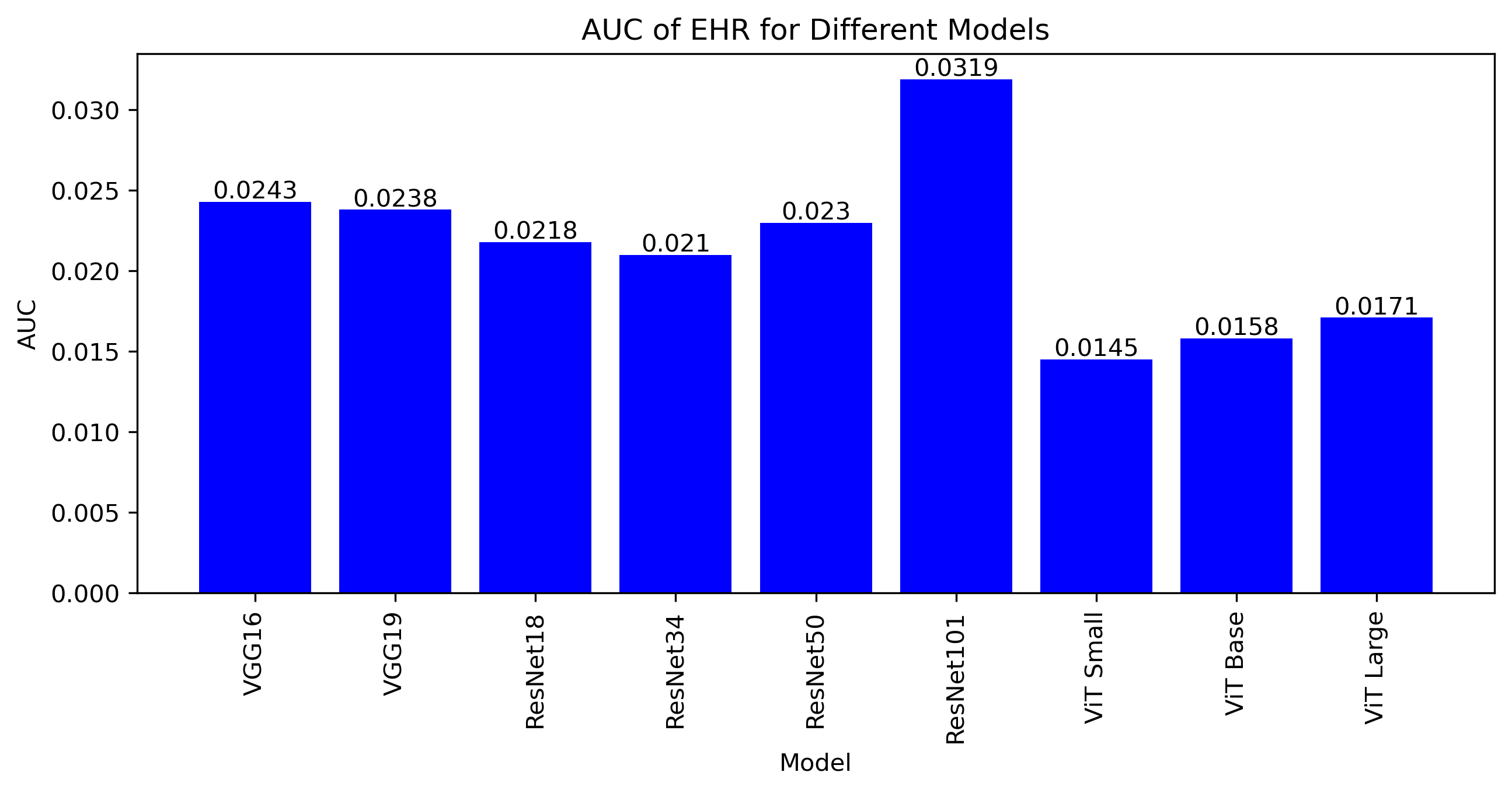}
\caption{\label{fig:ViT}EHR AUC results of of different models}
\end{figure}

\section{Discussion}
Deep learning architectures like VGG, ResNet, and ViT all achieved commendable diagnostic performance in the range of 0.84$\sim$0.88\%. As far as the experiments are concerned, deeper networks and the use of ViT do not contribute to better accuracy. Furthermore, with limited performance discrepancies between all tested models in pneumothorax diagnosis, the quality of Grad-CAM visualizations didn't necessarily correlate with the model's accuracy. The observed distinct behaviors of the Grad-CAM heatmaps may largely be due to the DL model's respective architecture types (pure CNN vs. Residual blocks vs. Transformer) and varying network sizes/depths. In our case study, while the Grad-CAM heatmaps of VGG16 and VGG19 accurately pinpointed areas of interest, VGG19 sometimes missed the pathological region, indicating that depth alone doesn't guarantee perfect feature localization. In addition, their activation maps often captured the upper chest areas, which may not be directly relevant for clinical interpretability. On the other hand, ResNet models, particularly ResNet18, 34, and 50, consistently highlighted relevant regions with a single cluster, albeit with slight deviations from the pneumothorax region in some cases. This could be attributed to the network's ability to focus on the most critical features through its residual connections. However, the much deeper ResNet101 model was prone to have two distinct areas in the Grad-CAM heatmaps. This was also noted by Seo et al. \cite{seo2021comparison} in their vertebral maturation stage classification with ResNets, likely due to the fact that deeper architectures can capture more intricate representations \cite{simonyan2014very}. In contrast, ViT models produced more dispersed Grad-CAM patterns compared to the included CNN ones. The inherent global contextual perception ability of Transformers might be responsible for this observation. As we only employed a single Chest X-ray dataset with a limited size, it's plausible that the ViTs could benefit from a larger dataset, potentially leading to better visualization outcomes \cite{steiner2021train}. Moreover, as the size of the ViT model increases, the proportion of irrelevant areas in the Grad-CAM visualizations also appears to increase, implying the interplay between model architecture and dataset size and the adverse cascading effects in deep Transformer models \cite{Zhou2021DeepViTTD}. In previous investigations for DL-based computer-assisted diagnosis \cite{lee2022advantages}\cite{seo2021comparison}, multiple network models of different designs and natures were often benchmarked together. However, to the best of our knowledge, a systematic investigation for the impact of network architecture types and network depths on Grad-CAM visualization, especially with quantitative assessments has not been conducted to date. As transparency is becoming increasingly important for the safety and adoptiblity of DL algorithms, the relevant insights are of great importance to the XAI community. 

The presented study has a few limitations. First,  a typical issue in medical deep learning is the lack of large, well-annotated datasets. To facilitate the training process, we employed DL models that were pre-trained using natural images and then fine-tuned using domain-specific data. As noted by Lee et al. \cite{lee2022advantages}, in comparison to training from scratch, model fine-tuning also has a better advantage to provide more sparse and specific visualization for the target regions in Grad-CAM heatmaps. Furthermore, additional data augmentation was also implemented to mitigate overfitting issues.  Second, as visual explanation of the DL algorithms is intended to allow easier incorporation of human participation, user studies to validate the quantitative metrics would be beneficial \cite{rong2022towards}. However, this requires more elaborate experimental design and inclusion of clinical experts, and will be included in our future studies. Lastly, we utilized pneumothorax diagnosis in chest X-ray as a case study to investigate the impact of DL model architectures on Grad-CAM visualization. It is possible that the observed trend may be application-specific. To confirm this, we will explore different datasets with varying disease types and imaging contrasts in the future.

\section{Conclusion}
In this study, we performed a comprehensive assessment of popular DL architectures, including VGG, ResNet and ViT, and their variants of different sizes for pneumothorax diagnosis in chest X-ray, and investigated the impacts of network depths and architecture types on visual explanation provided by Grad-CAM. While the accuracy for pneumothorax vs. healthy classification is similar across different models, the CNN models offer better specificity and accuracy than the ViTs when comparing the resulting heatmaps from Grad-CAM. Furthermore, the network size can affect both the model accuracy and Grad-CAM outcomes, with the two factors not necessarily in synch with each other. We hope the insights from our study can help better inform future explainable AI research, and we will further confirm the observations with more extensive studies involving more diverse datasets and DL models in the near future.

%
%
%
%

\end{document}